\documentclass[aps,prc,twocolumn,superscriptaddress,showpacs]{revtex4}

\usepackage{graphicx}

\begin{document}


\title{Gamow-Teller transitions to $^{64}$Cu measured using the $^{64}$Zn(t,$^{3}$He) reaction}

\author{G.W. Hitt}
\affiliation{{National Superconducting Cyclotron Laboratory, Michigan State University, East Lansing, MI 48824-1321, USA}}
\affiliation{{Department of Physics and Astronomy, Michigan State University, East Lansing, MI 48824, USA}}
\affiliation{{Joint Institute for Nuclear Astrophysics, Michigan State University, East Lansing, MI 48824, USA}}

\author{R.G.T. Zegers}
\email{zegers@nscl.msu.edu}
\affiliation{{National Superconducting Cyclotron Laboratory, Michigan State University, East Lansing, MI 48824-1321, USA}}
\affiliation{{Department of Physics and Astronomy, Michigan State University, East Lansing, MI 48824, USA}}
\affiliation{{Joint Institute for Nuclear Astrophysics, Michigan State University, East Lansing, MI 48824, USA}}

\author{Sam M. Austin}
\affiliation{{National Superconducting Cyclotron Laboratory, Michigan State University, East Lansing, MI 48824-1321, USA}}
\affiliation{{Joint Institute for Nuclear Astrophysics, Michigan State University, East Lansing, MI 48824, USA}}

\author{D. Bazin}
\affiliation{{National Superconducting Cyclotron Laboratory, Michigan State University, East Lansing, MI 48824-1321, USA}}

\author{A. Gade}
\affiliation{{National Superconducting Cyclotron Laboratory, Michigan State University, East Lansing, MI 48824-1321, USA}}
\affiliation{{Department of Physics and Astronomy, Michigan State University, East Lansing, MI 48824, USA}}

\author{D. Galaviz}
\altaffiliation[Present address: ]{Centro de Fisica Nuclear da Universidade de Lisboa, 1649-003, Lisboa, Portugal}
\affiliation{{National Superconducting Cyclotron Laboratory, Michigan State University, East Lansing, MI 48824-1321, USA}}
\affiliation{{Joint Institute for Nuclear Astrophysics, Michigan State University, East Lansing, MI 48824, USA}}

\author{C.J. Guess}
\affiliation{{National Superconducting Cyclotron Laboratory, Michigan State University, East Lansing, MI 48824-1321, USA}}
\affiliation{{Department of Physics and Astronomy, Michigan State University, East Lansing, MI 48824, USA}}
\affiliation{{Joint Institute for Nuclear Astrophysics, Michigan State University, East Lansing, MI 48824, USA}}

\author{M. Horoi}
\affiliation{{Department of Physics, Central Michigan University, Mount Pleasant, MI 48859, USA}}

\author{M.E. Howard}
\affiliation{{Department of Physics, The Ohio State University, Columbus, OH 43210, USA}}
\affiliation{{Joint Institute for Nuclear Astrophysics, Michigan State University, East Lansing, MI 48824, USA}}

\author{Y. Shimbara}
\altaffiliation[Present address: ]{Graduate School of Science and Technology, Niigata University, Niigata 950-2181, Japan}
\affiliation{{National Superconducting Cyclotron Laboratory, Michigan State University, East Lansing, MI 48824-1321, USA}}
\affiliation{{Joint Institute for Nuclear Astrophysics, Michigan State University, East Lansing, MI 48824, USA}}

\author{E.E. Smith}
\affiliation{{Department of Physics, The Ohio State University, Columbus, OH 43210, USA}}
\affiliation{{Joint Institute for Nuclear Astrophysics, Michigan State University, East Lansing, MI 48824, USA}}

\author{C. Tur}
\affiliation{{National Superconducting Cyclotron Laboratory, Michigan State University, East Lansing, MI 48824-1321, USA}}
\affiliation{{Joint Institute for Nuclear Astrophysics, Michigan State University, East Lansing, MI 48824, USA}}

\date{\today}

\begin{abstract}

The $^{64}$Zn($t$,$^{3}$He) reaction has been studied using a secondary triton beam of 115 MeV/nucleon to extract the Gamow-Teller transition-strength distribution to $^{64}$Cu. The results were compared with shell-model calculations using the $pf$-shell effective interactions KB3G and GXPF1A and with existing data from the $^{64}$Zn($d$,$^{2}$He) reaction. Whereas the experimental results exhibited good consistency, neither of the theoretical predictions managed to reproduce the data. The implications for electron-capture rates during late stellar evolution were investigated. The rates based on the theoretical strength distributions are lower by factors of 3.5-5 compared to the rates based on experimental strength distributions.
\end{abstract}


\maketitle

\section{\label{sec:intro}Introduction}

Weak reaction rates play an important role in late stellar evolution \cite{lang2003}. In particular, nuclear electron captures (EC) affect the pre-explosion evolution of core-collapse supernovae \cite{beth1979}, where the high Fermi energy of degenerate electrons lift Q-value restrictions on EC that are present at terrestrial electron densities. The EC reactions modify properties of the stellar core by neutronizing nuclear matter and reducing the electron abundance ($Y_{e}$). As a consequence, pre-collapse dynamics are modified due to the reduced outward pressure. In addition, energy and entropy are reduced by neutrino emission associated with the weak transitions \cite{hix2003}. Fuller, Fowler and Newman (FFN) first showed that EC in core-collapse supernovae are predominantly allowed Gamow-Teller (GT) transitions in $pf$- and $sdg$-shell nuclei \cite{full1980,full1982a,full1982b,full1985}. They provided the first parameterized EC rates for $pf$-shell nuclei, treating nuclear excitations in an independent-particle model (IPM) up to $A=60$ and assigning a flat GT response to heavier nuclei.

It has since then been understood that residual interactions between constituent nucleons move the centroid of the GT strength (B(GT); here defined such that B(GT)=3 for the $\beta$-decay of a free neutron) distribution and spread it over many more states in the daughter than assumed in the IPM used by FFN \cite{brow1988}. Shell-model calculations \cite{caur1999,lang2000} employing effective interactions \cite{pove2001,honm2005} with parameters that are fitted to experimental data predict significantly reduced low-lying GT strengths and hence reduced EC rates compared to the estimates by FFN. Core-collapse models using these revised rates \cite{hege2001a,hix2003} indicate that the pre-explosion evolution is strongly impacted by the reduced EC rates. It is, therefore, crucial that the theoretical models that aim to describe GT strength distributions are tested against data. The features of the GT strength distributions at low excitation energies are particularly important \cite{lang2003}. At any given time prior to the collapse, a varying ensemble of nuclei plays a significant role \cite{hege2001b,hix2003}. Therefore, it is important to systematically test theory against experiment over a broad mass range so that the effective interactions can be improved.

Investigations of charge-exchange (CE) reactions at incident energies $\gtrsim$ 100 MeV/nucleon \cite{oste1992,hara2001} are the preferred method for this purpose since, in contrast to $\beta$-decay studies, the full excitation-energy region of interest for the astrophysical applications can be probed. In the $\Delta T_{z}=+1$ direction, which is of direct relevance for electron-capture, the ($n$,$p$) (see e.g. \cite{alfo1993,elka1994}), ($d$,$^{2}$He) (see e.g. \cite{hage2004,baum2005,grew2008}) and ($t$,$^{3}$He) \cite{cole2006} probes have been employed. Experimental studies of $\Delta T_{z}=-1$ probes, such as ($p$,$n$) and ($^{3}$He,$t$) have also been used to extract GT strength distributions of relevance for EC, under the assumption of isospin symmetry \cite{fuji2005,anan2008}. The foundation for the experimental work is the empirical proportionality between B(GT) and differential cross section for GT transitions at vanishing momentum transfer, discussed in more detail below.

In this work, we employed the $^{64}$Zn($t$,$^{3}$He) charge-exchange reaction at 115 MeV/nucleon incident triton energy to extract the GT transition strength to $^{64}$Cu. The experimental results were compared to shell-model calculations employing the effective interactions KB3G \cite{pove2001} and GXPF1A \cite{honm2005}. EC rates on the $^{64}$Zn groundstate were calculated as a function of stellar temperature and density to investigate the impact of the differences between the measured and theoretical GT strength distributions. We also compare our results with those of a recent $^{64}$Zn($d$,$^{2}$He) experiment, which was aimed at extracting information on GT strengths in relation to double-$\beta$ decay transition matrix elements \cite{grew2008}.

\section{\label{sec:exper}Experiment}
A secondary triton beam was produced by fast fragmentation of a 150-MeV/nucleon $^{16}$O$^{8+}$ primary beam on a thick (3500 mg/cm$^{3}$) $^{9}$Be production target at National Superconducting Cyclotron Laboratory's (NSCL) Coupled Cyclotron Facility (CCF) \cite{ccf}. The A1900 Fragment Separator \cite{a1900} was set to a magnetic rigidity $B\rho =4.8$ Tm, corresponding to 115 MeV/nucleon triton energy. A slit at the intermediate image of the A1900 was used to limit the momentum acceptance to $dp/p=\pm 0.21$\%, corresponding to $\sim$3 MeV triton kinetic energy spread. The isotopic purity of the triton beam was 85\%, with $^{6}$He being the principle contaminant, and the average intensity at the reaction target was $\sim 3\times 10^{6}$ s$^{-1}$. Details of the production of the secondary triton beam can be found in Refs. \cite{hitt2006,howa2008}.

A 9.84-mg/cm$^{2}$ thick $^{64}$Zn foil, with an isotopic purity of 99.6\% was placed at the pivot-point of the S800 spectrograph \cite{s800}. The analysis beam line to the target was operated in momentum dispersion-matching mode \cite{fuji2002} to optimize the energy resolution.  The S800 spectrograph was set at 0$^{\circ}$ laboratory scattering angle. Although it is at present possible to monitor the absolute triton beam intensity at the target with good precision (uncertainty $\pm 5$\%), at the time of the $^{64}$Zn($t$,$^{3}$He) experiment systematic errors were significantly higher. Therefore, using the $^{12}$C($t$,$^{3}$He) reaction, the excitation of the strong GT transition from the $^{12}$C 0$^{+}$ groundstate to the $^{12}$B 1$^{+}$ groundstate with known cross section \cite{cole2006,perd2009} served to calibrate the absolute beam intensity. A deuterated polyethylene foil (CD$_{2}$) with a thickness of 9.10 mg/cm$^{2}$ was used for that purpose. Minor variations of the beam intensity during $^{64}$Zn runs due to fluctuations in the $^{16}$O primary beam intensity were monitored using a Faraday bar placed in the first dipole magnet of the A1900.

\begin{figure}
\begin{center}
\includegraphics[width=0.49\textwidth]{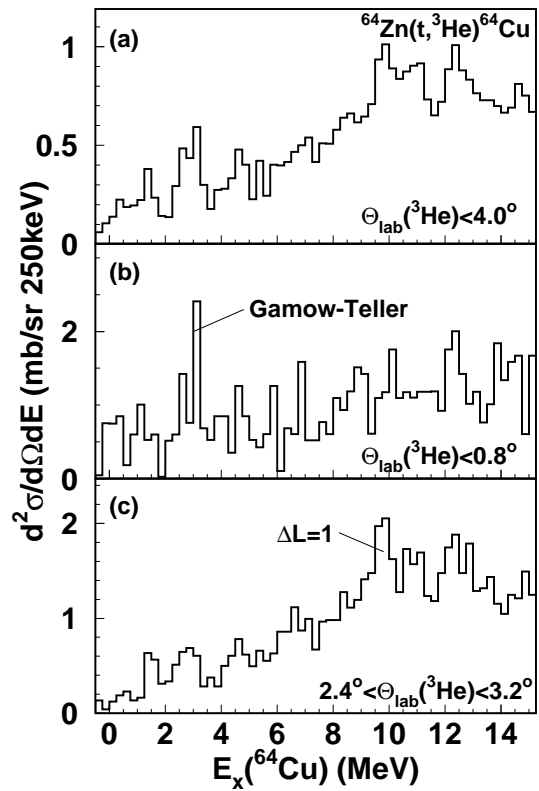}
\caption{Differential cross section of the $^{64}$Zn(t,$^{3}$He)$^{64}$Cu CE reaction, plotted as a function of excitation energy ($E_x$) in $^{64}$Cu. (a) The spectrum obtained for the full solid angle covered in the experiment. (b) The spectrum gated on forward angles $\Theta_{\rm lab}$($^{3}$He)$<0.8^{\circ}$. (c) The spectrum gated at backward angles $2.4^{\circ}<\Theta_{\rm lab}$($^{3}$He)$<3.2^{\circ}$ \label{fig1}.}
\end{center}
\end{figure}

$^{3}$He ejectiles were momentum analyzed in the S800 focal plane \cite{yurk1999} using two cathode-readout drift chambers (CRDCs) as tracking detectors and two plastic scintillators ($E1$ and $E2$) to measure energy loss and event time-of-flight (TOF). The $E1$ signal was the data-acquisition master trigger and TOF start. The CCF radio frequency (RF) signal was the TOF stop. Ion energy loss in $E1$ and event TOF allowed for unambiguous identification of $^{3}$He events in the S800 focal plane. A 5$^{th}$-order inverse ion-optical transfer matrix, calculated based on measured magnetic field maps with the code \textsc{COSY Infinity} \cite{cosy}, was used to reconstruct the $^{3}$He momentum and scattering angle ($\Theta_{\rm lab}$($^{3}$He)). The excitation energy in $^{64}$Cu ($E_{x}$($^{64}$Cu)) was calculated in a missing-mass calculation. In Fig. \ref{fig1}(a), the excitation energy spectrum up to 15 MeV for the $^{64}$Zn($t$,$^{3}$He) reaction is displayed. The excitation energy resolution was 280 keV (FWHM). Events with a scattering angle of up to $4^{\circ}$  were included in the analysis. The scattering angle resolution was 10 mrad (FWHM).

GT transitions are associated with spin-transfer ($\Delta S = 1$) and zero units of angular-momentum transfer ($\Delta L = 0$). In the case of the $^{64}$Zn target (groundstate $J^{\pi}=0^{+}$), $1^{+}$ states are populated in $^{64}$Cu.
The GT differential cross section peaks at $0^{\circ}$, which separates them from transitions associated with single or multiple units of angular-momentum transfer. Since GT transitions are in general not separated in excitation energy from other transitions, a multipole decomposition analysis (MDA) is required to isolate the GT contribution to the excitation-energy spectrum. In addition, $0^{+} \rightarrow 1^{+}$ transitions contain a $\Delta L=2$, $\Delta S=1$ component. Near $0^{\circ}$, it is usually small compared to the $\Delta L=0$, $\Delta S=1$ contribution, but the two contributions can only be separated through a MDA. The MDA performed for this work is described in more detail in section \ref{sec:mda}.

Using the Eikonal approximation, Taddeucci \textit{et al.} \cite{tadd1987} showed that for $E_{\rm beam}\gtrsim 100$ MeV/nucleon, the differential cross section at vanishing momentum transfer ($q=0$) is proportional to B(GT):
\begin{equation}
\frac{d\sigma}{d\Omega}\Bigg\vert_{q\rightarrow0}=\hat{\sigma}_{\rm GT}B({\rm GT}),
\label{eq:csbgt}
\end{equation}
where $\hat{\sigma}_{\rm GT}$ is referred to as the unit cross section. This relationship, first studied for the ($p$,$n$) reaction, has since been employed for a variety of CE probes. Because of the similarity of the ($^{3}$He,$t$) and ($t$,$^{3}$He) probes, the most relevant for the present work is the recently empirically established \cite{zege2007} relationship between the unit cross section and target mass number for the ($^{3}$He,t) reaction at 140 MeV/nucleon, as discussed in detail in Section \ref{sec:unitcs}.

The proportionality expressed in Eq. \ref{eq:csbgt} is not perfect. The main source of proportionality breaking is the interference between $\Delta L=0$, $\Delta S=1$ and $\Delta L=2$, $\Delta S=1$ amplitudes due to the non-central tensor component of the projectile-target interaction \cite{zege2006,cole2006}. The effects are difficult to quantify experimentally, but can be estimated theoretically, as is done in Section \ref{sec:error}.

\subsection{\label{sec:mda}Multipole Decomposition Analysis (MDA)}

In Fig. \ref{fig1}, the experimental spectra over the full opening angle covered (Fig. \ref{fig1}(a)) and gated on scattering angles at or below $0.8^{\circ}$ (Fig. \ref{fig1}(b)) and between $2.4^{\circ}$ and $3.2^{\circ}$ (Fig. \ref{fig1}(c)) are shown. The energy bin size is 250 keV, which is close to the experimental energy resolution. For states below the threshold for decay by particle emission (the binding energies for protons and neutrons are 7.2 MeV and 7.9 MeV, respectively), the line-width is much less than the experimental resolution and below 4$^{\circ}$ scattering angle most of the yield thus falls within one or two energy bins.  States associated with $\Delta L=0$ peak at forward scattering angle and are thus relatively strong in Fig. \ref{fig1}(b) and weak in Fig. \ref{fig1}(c). The peak seen in the bin at an excitation energy of 3.125 MeV is the clearest example.
In contrast, the structure seen in the spectra at $\sim 10-12$ MeV has a relatively large cross section at backward scattering angles (Fig. \ref{fig1}(c)) but is not easily identified at forward angles (Fig. \ref{fig1}(b)). This is a characteristic of dipole transitions ($\Delta L=1$), which peak at $\Theta_{\rm lab}$($^{3}$He)=$2-3^{\circ}$ and have a local minimum at $\Theta_{\rm lab}$($^{3}$He)=$0^{\circ}$. Differential cross sections for transitions of higher angular-momentum transfers are relatively flat at forward angles, with a less-pronounced maximum at $\Theta_{\rm lab}$($^{3}$He)=$3-4^{\circ}$. It is further noted that, apart from the GT transitions, the only other significant $\Delta L=0$ contributions to the excitation energy spectrum in the $T_{z}=+1$ direction stem from isovector (spin-flip) giant monopole resonances. They have negligible cross sections below $E_{x}\approx10$ MeV \cite{auer1984}. In this work, the B(GT) in $^{64}$Cu is extracted only up to 6 MeV. Consequently, the assignment of $\Delta L=0$ contributions to GT transitions is unambiguous.
\begin{figure}
\begin{center}
\includegraphics[width=0.49\textwidth]{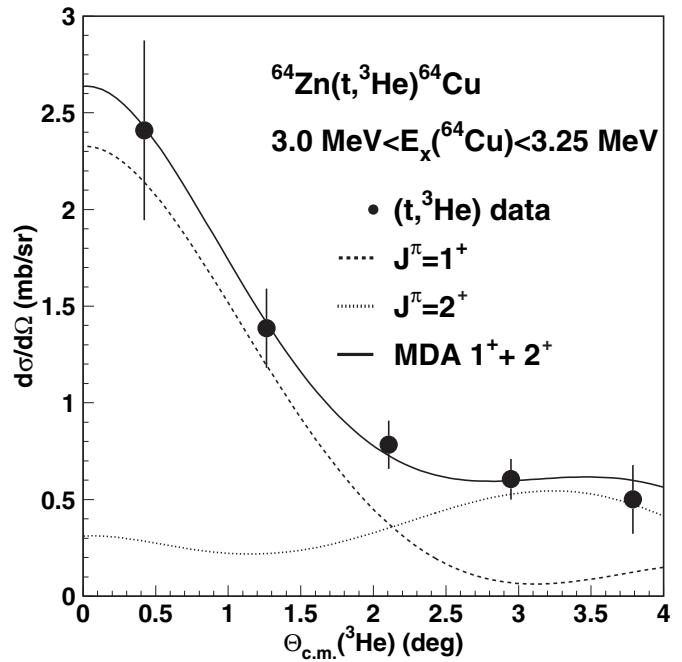}
\caption{Differential cross section from the bin centered at 3.125 MeV seen in Fig. \ref{fig1}(b).\label{fig2}}
\end{center}
\end{figure}
To extract the magnitude of the $\Delta L=0$ cross section in each 250-keV wide bin, a MDA was performed. In Fig. \ref{fig2}, an example of the procedure is shown for the bin centered at 3.125 MeV. The $^{3}$He angular distributions were generated in the Distorted Wave Born Approximation (DWBA) with the DWBA code FOLD \cite{dwhi}; transitions involving $\Delta L=0$, $\Delta L=1$ and $\Delta L=2$ to $^{64}$Cu were considered. Transitions with $\Delta L>2$ are suppressed at these beam energies and their angular distributions below 4$^{\circ}$ scattering angle are quite similar to the $L=1,2$ shapes already used in the MDA. The effective NN interaction by Love and Franey at 140 MeV \cite{love1981,fran1985} was double-folded over the transition densities of the $t$-$^{3}$He and $^{64}$Zn-$^{64}$Cu systems to generate the form factor. A short-range approximation \cite{love1981} was used to describe exchange processes.
Binding energies of single-particle wave functions in $^{64}$Zn and $^{64}$Cu were calculated in the code \textsc{OXBASH} \cite{OXBA} using the Skyrme SK20 interaction \cite{BRO98}. The $^{3}$He and $^{3}$H densities were obtained from Variational Monte-Carlo results \cite{WIR05}. One-body transition densities for GT transitions were generated in \textsc{NuShellX} \cite{NSX} using the GXPF1A \cite{honm2005} interaction.

The shape of the $\Delta L=0$ angular distributions do not vary significantly for transitions to different states. For transitions associated with $\Delta L=1$ or 2, one-body transitions densities were generated in a normal-mode formalism using the code \textsc{NORMOD} \cite{NOR}.
Optical-potential parameters for both entrance and exit channels were taken or deduced from $^{3}$He elastic-scattering data on $^{58}$Ni \cite{kami2003}. The $^{3}$He-$^{64}$Cu optical-potential parameters were -35.16 MeV (-44.43 MeV), 1.320 fm (1.021 fm), and 0.840 fm (1.018 fm) for real (imaginary) Woods-Saxon well-depth, radius and diffuseness, respectively. For the $t$-$^{64}$Zn system, the above well-depths were scaled by a factor 0.85, following the procedure in Ref. \cite{werf1989}.

Pair-wise combinations of calculated angular distributions associated with different units of angular-momentum transfer were used to fit the experimental angular distribution for each excitation-energy bin. Only the normalizations of the calculated angular distributions were taken as free fit parameters. Due to the limited angular coverage, fits with more than two components did not improve the reduced $\chi^{2}$ value of the fit, nor change the cross section of the $\Delta L=0$ contribution by more than a fraction of the statistical errors. The fit with the combination of angular distributions minimizing the reduced $\chi^{2}$ value was chosen as the best description of the data. The angular distribution of the bin centered around 3.125 MeV shown in Fig. \ref{fig2} was best described by a strong $J^{\pi}=1^{+}$ component in combination with a weaker $J^{\pi}=2^{+}$ contribution. Statistically significant $\Delta L=0$ contributions to the spectrum were found up to an excitation energy of 6 MeV. In each energy bin, the fitted $\Delta L=0$ curve was used to determine the cross section at 0$^{\circ}$. The fit error was used to determine the uncertainty in the extracted 0$^{\circ}$ cross section. In some bins no significant $\Delta L=0$ contribution was found. Although it cannot be excluded that small $\Delta L=0$ contributions are present in these bins due to the limited statistics, we report no Gamow-Teller strength for such cases.

The 0$^{\circ}$ cross section for a transition at $Q<0$ must be extrapolated to $Q=0$ to reach the zero linear-momentum transfer ($q=0$) limit required for application of Eq. \ref{eq:csbgt}. The extrapolation factor was determined using the relation:
\begin{equation}
\frac{d\sigma}{d\Omega}\Bigg\vert_{q\rightarrow0}=\bigg[\frac{\frac{d\sigma}{d\Omega}(Q=0,0^{\circ})}{\frac{d\sigma}{d\Omega}(Q,0^{\circ})}\bigg]_{\rm DWBA} \bigg[\frac{d\sigma}{d\Omega}(Q,0^{\circ})\bigg]_{\rm exp},
\label{eq:qzero}
\end{equation}
where the subscript ``${\rm DWBA}$'' denotes calculated cross sections in DWBA and the subscript  ``${\rm exp}$'' denotes the measured cross section. The DWBA calculations were performed for 100 reaction $Q$-values so that the ratio of DWBA cross sections in Eq. \ref{eq:qzero} could be established as a smooth function of excitation energy.

\subsection{\label{sec:unitcs}Application of the ($^{3}$He,t) Unit Cross Section}

From the extracted cross sections at $q=0$, absolute B(GT)s were determined by dividing the measured cross section by the unit cross section $\hat{\sigma}_{GT}$. The unit cross section is often calibrated directly using a transition for which the B(GT) is known from a $\beta$-decay measurement. In principle, this is possible for the case of $^{64}$Zn$\rightarrow$$^{64}$Cu as well, since the log$ft$ for the groundstate-to-groundstate transition is known \cite{sing2007}. The measured value is log$ft$=5.301$\pm$0.006, which, following the formalism of Ref. \cite{BRO93}, corresponds to a B(GT) for the $\beta$-decay of $^{64}$Cu of $0.0192\pm0.0004$. Taking into account that $B(\textrm{GT})_{a\rightarrow b}=(2J_{b}+1)/(2J_{a}+1)B(\textrm{GT})_{b\rightarrow a}$, the B(GT) for the $^{64}$Zn$\rightarrow^{64}$Cu transition is $0.058\pm0.001$. This direct calibration of the unit cross section was used in the analysis of the $^{64}$Zn($d$,$^{2}$He) experiment \cite{grew2008}. However, due to the limited energy resolution and statistics in the present ($t$,$^{3}$He) data set, the relatively weak groundstate transition could not be resolved from the $1^{+}$ excited state at 344 keV \cite{sing2007}. Moreover, as discussed in the next section, the calibration of the unit cross section using a $\beta$-decay strength to a weak state can lead to significant errors if the proportionality breaking is large for the particular transition used due to effects of the tensor interaction (see e.g. the case $^{58}$Ni \cite{cole2006}).

As an alternative method, we use the empirically established mass-dependent trend line for the unit cross section of the ($^{3}$He,$t$) probe \cite{zege2007}:
\begin{equation}
\hat{\sigma}_{\rm GT}=109 A^{-0.65},
\label{eq:unitcs}
\end{equation}
where $A$ is the target mass number. It was recently found that the unit cross sections for the ($t$,$^{3}$He) reaction in the case of $A$=12,13 and 26 were consistent within error margins to that of the ($^{3}$He,$t$) reaction \cite{perd2009}, in spite of the minor difference in beam energy (115 MeV/nucleon for ($t$,$^{3}$He) versus 140 MeV/nucleon for ($^{3}$He,$t$)). Although future ($t$,$^{3}$He) experiments over a wider mass range are important to further establish the consistency of the unit cross sections between the two inverse reactions, it seems likely that the close correspondence will be maintained for higher mass numbers.

\begin{figure}
\begin{center}
\includegraphics[width=0.49\textwidth]{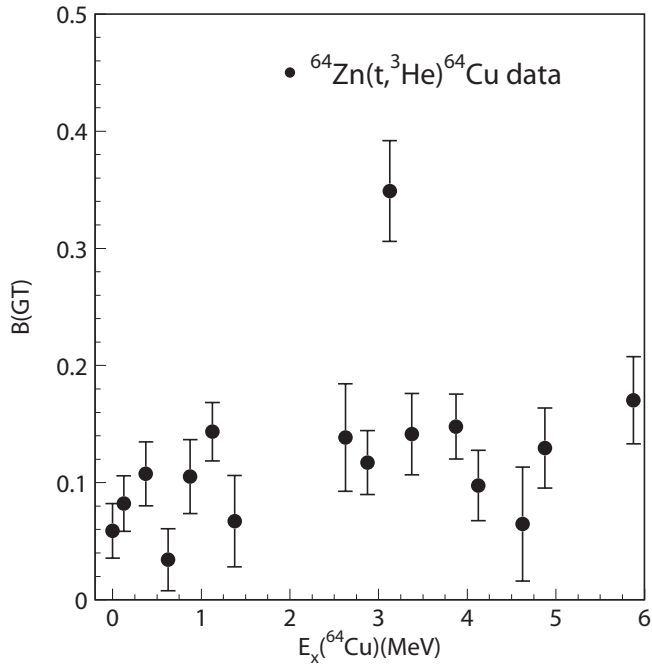}
\caption{The GT strength in $^{64}$Cu, extracted up to 6 MeV in excitation energy.\label{fig3}}
\end{center}
\end{figure}

In summary, Eqs. \ref{eq:csbgt}, \ref{eq:qzero}, and \ref{eq:unitcs} were used to calculate the absolute B(GT) of transitions to states in $^{64}$Cu based on the extracted $\Delta L=0$ partial cross sections. Fig. \ref{fig3} shows the resulting B(GT) distribution in $^{64}$Cu. The transition to the 3.125 MeV bin was identified as having the highest B(GT).
The B(GT) values in the first two 250-keV wide bins were adjusted to account for the fact that the GT strength for the groundstate transition is known. Hence, the B(GT) value in the first bin was adjusted to the value obtained from $\beta$-decay. The remaining strength observed experimentally in that first excitation-energy bin was shifted to the second excitation-energy bin under the assumption that, due to the limited resolution, some of the higher-lying strength was detected at too low an excitation energy.

The errors presented in Fig. \ref{fig3} are dominated by the statistical error, with a small contribution from uncertainties due to the assumptions used in the fitting procedure. There is an  uncertainty in the overall scaling  of about 10\% due the relative normalization to the $^{12}$C($t$,$^{3}$He)$^{12}$B(1$^{+}$) reaction and the application of the unit cross section as calculated in Eq. \ref{eq:unitcs}.

\subsection{\label{sec:error}Analysis of Systematic Errors}
As mentioned above, the proportionality between B(GT) and the cross section at $q=0$ is not perfect, due to interference between $\Delta L=0$, $\Delta S=1$ and $\Delta L=2$, $\Delta S=1$ amplitudes. The latter amplitude is mediated through the tensor component of the effective NN interaction. In general, it is difficult to evaluate these errors on a transition-by-transition basis, since it requires an accurate knowledge of the contributions from individual $1p-1h$ components to the transition density \cite{cole2006}. However, as shown in Ref. \cite{zege2006}, a theoretical study of the proportionality breaking for many shell-model states can reveal the associated uncertainties as a function of B(GT) and the same procedure was carried out for the $^{64}$Zn($t$,$^{3}$He) reaction and is described in this section.

\begin{figure}
\begin{center}
\includegraphics[width=0.45\textwidth]{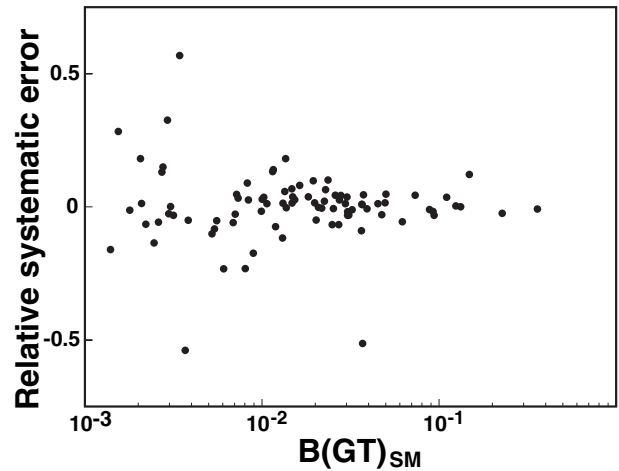}
\caption{A theoretical estimation of the relative systematic error in the B(GT$_{+}$) extraction using Eq. \ref{eq:csbgt}, plotted as a function of the shell-model strength.\label{fig6}}
\end{center}
\end{figure}

To evaluate the systematic error due to the interference, the one-body transition densities of 100 GT shell-model transitions generated with the GXPF1A \cite{honm2005} interaction (all states predicted below about 6 MeV) were used to calculate differential cross section in DWBA. From there on, the theoretical cross sections were treated as if they were data and B(GT)s were extracted as described above. Since the extraction assumes a common unit cross section for all transitions, the deviation between the extracted strengths ($B(\textrm{GT})_{\textrm{\scriptsize DWBA}}$) and the values calculated in the shell model ($B(\textrm{GT})_{\textrm{\scriptsize SM}}$) provides a measure for the amount of proportionality breaking due to the interference and other effects included in the DWBA calculations.
A relative systematic error was defined as:
\begin{equation}
{\rm Rel.sys.err.}=\frac{B({\rm GT})_{\rm DWBA}-B({\rm GT})_{\rm SM}}{B({\rm GT})_{\rm SM}}.
\label{eq:syserr}
\end{equation}

Fig. \ref{fig6} shows the relative systematic error for the 100 $^{64}$Zn($t$,$^{3}$He)$^{64}$Cu($1^{+}$) transitions generated, plotted as a function of the B(GT) determined in the shell-model. On average, the relative systematic error increases with decreasing B(GT). An approximate relationship between the width of the error distribution and the B(GT) was extracted from these points:
\begin{equation}
\sigma_{\textrm{rel.sys.err.}}\approx 0.03-0.033\times {\rm ln}(B({\rm GT})).
\label{eq:errbgt}
\end{equation}
It is noted that this relationship is very similar to that found for the case of $^{26}$Mg($^{3}$He,$t$) \cite{zege2006}, in spite of the fact that the shell-model spaces and interactions employed are very different.

The procedure for estimating the breaking of the unit cross section in the $^{64}$Zn case was performed a second time after switching off the tensor contributions to the effective NN interaction \cite{love1981,fran1985}. The relative systematic error as defined in Eq. \ref{eq:syserr} became much smaller; $\sigma_{\textrm{rel.sys.err.}}<5$\%, even for the weakest transitions. These remaining errors are due to effects such as exchange contributions and small differences in the shape of the transition densities for different GT transitions, but are much smaller than the effects due to the tensor interaction.

There are outliers in the relative systematic error distribution. For example, for the state seen in Fig. \ref{fig6} at $B(\rm GT)_{\rm SM}\approx 0.03$, which corresponds to the first excited state in the theoretical spectrum, the proportionality is broken by $\approx 50\%$. If such a strong breaking were to happen for the groundstate transition and that transition were to be used for the overall calibration of the unit cross section, instead of the present procedure, the same large error would be introduced for all transitions. Applying the ($^{3}$He,$t$) unit cross section as described in Eq. \ref{eq:unitcs} to ($t$,$^{3}$He) cross section data to extract B(GT) guards one against this potentially large systematic error.

\subsection{\label{sec:dhecom}Consistency with the ($d$,$^{2}$He)  Reaction}

\begin{figure}
\begin{center}
\includegraphics[width=0.48\textwidth]{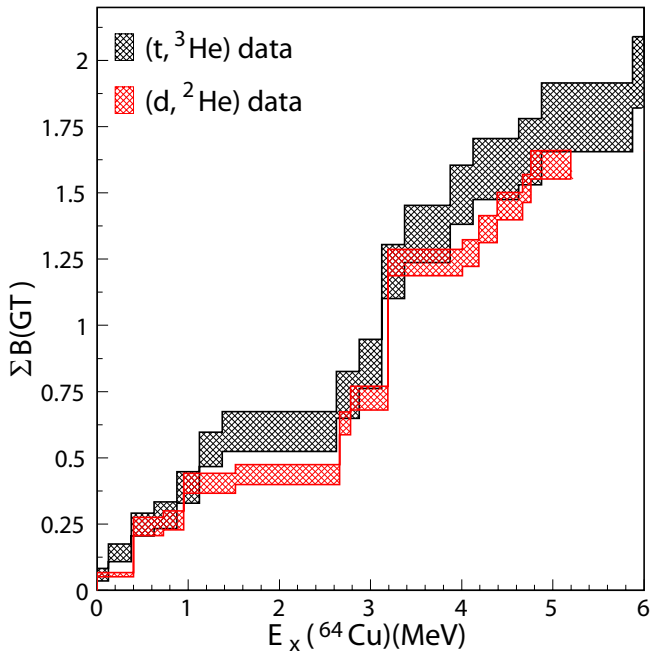}
\caption{(color online) Running sum of B(GT) in $^{64}$Cu, plotted as a function of excitation energy, for B(GT) extracted with the ($t$,$^{3}$He) reaction (black, thick cross hatches) and with the ($d$,$^{2}$He) reaction (red, thin cross hatches) \cite{grew2008}.\label{fig4}}
\end{center}
\end{figure}
In Fig. \ref{fig4} the running sum of B(GT) as a function of the excitation energy in $^{64}$Cu is compared with that of the $^{64}$Zn($d$,$^{2}$He) measurement of Grewe \textit{et al.} \cite{grew2008}. The width of the bands represent the cumulative errors in the summations. Note that in Ref. \cite{grew2008}, B(GT)s were extracted only up to 5 MeV. The main features of the two distributions are consistent. A difference of $\approx 10$\% in the overall strength up to an excitation energy of 5 MeV is found, but the deviations are within error margins, which, moreover, do not include errors in the overall B(GT) normalization for either data set. The summed B(GT)s up to an excitation energy of 5 MeV are $\sum B({\rm GT})=1.78\pm 0.13$ in the present work and $\sum B({\rm GT})=1.61\pm 0.05$ for the results from the ($d$,$^{2}$He) experiment.

The reactions and methods employed to extract the B(GT) distribution were quite different for the present work and that of Grewe \textit{et al.} \cite{grew2008}. Here, the assumption was made that the unit cross section from ($^{3}$He,$t$) could be employed and the analysis of the ($d$,$^{2}$He) data relied on the fact that the proportionality between differential cross section at $q=0$ and B(GT) was not strongly broken for the groundstate transition that was used to calibrate the unit cross section. The fact that the results are consistent gives confidence that both assumptions were reasonable, on the level of $\approx 10$\%.

\section{\label{sec:theory}Comparison with shell-model calculations}

The goal of the present work is to test the theoretical predictions for weak-transition strengths that are used to generate electron-capture rates of relevance for stellar evolution. In this section, we compare theoretical and experimental GT strength distributions. In the next section, the implications for the electron-capture rates under stellar conditions are explored.

The shell-model has provided the best description of low-lying GT strength distributions \cite{caur1999,lang2000}. Here, our experimental results are compared with shell-model calculations using the $pf$-shell effective interactions GXPF1A \cite{honm2005} and KB3G \cite{pove2001}. The parameters of the GXPF1A interaction have been fitted to reproduce the experimental excitation energies and masses (but not GT strengths) over a wide range of $pf$-shell nuclei including A=64. KB3G is a slightly modified version of the KBF interaction \cite{caur1999}. Their parameters have been deduced from experimental data in the lower $pf$-shell only. The KBF interaction has been used to generate weak transition rate tables \cite{lang2001a} that are commonly used in astrophysical simulations.

\begin{figure}
\begin{center}
\includegraphics[width=0.49\textwidth]{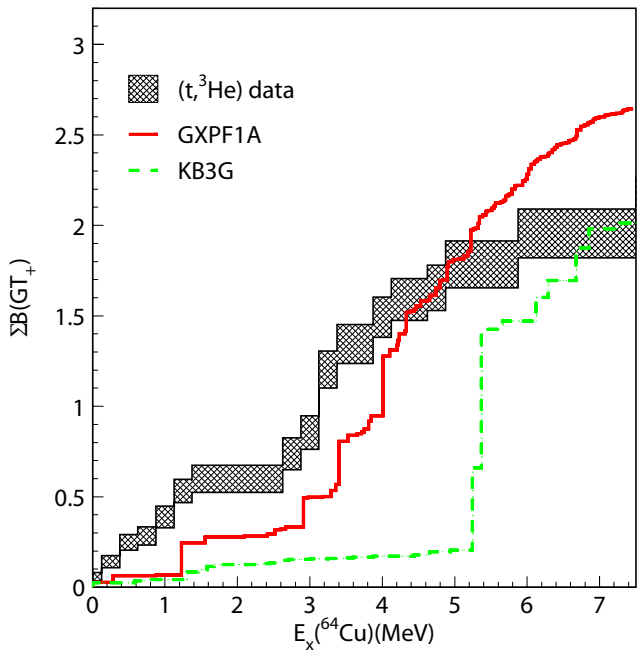}
\caption{(color online) Running sum of B(GT) in $^{64}$Cu, plotted as a function of excitation energy, as measured with the ($t$,$^{3}$He) reaction (black, cross hatched), and calculated with the shell-model using GXPF1A \cite{honm2005} (red, solid line) and KB3G \cite{pove2001} (green, dashed line) effective interactions. \label{fig5}}
\end{center}
\end{figure}

The shell-model calculations were performed using the code \textsc{NuShellX} \cite{NSX}, which was slightly modified to provide a large number of states of a given spin. Figure \ref{fig5} shows the running sum of B(GT) in $^{64}$Cu up to an excitation energy of 7.5 MeV. Results from the present experimental work and from the two theoretical calculations using the GXPF1A \cite{honm2005} and KB3G \cite{pove2001} interactions are shown. The theoretical calculations have been scaled by a $pf$-shell specific quenching factor of $(0.74)^{2}$ \cite{mart1996} to account for effects not included in the model space.

Neither calculation reproduces the experimental strength distribution. The distribution for GXPF1A is closer to the data, but pushes the strength up too high in excitation energy. The same, but more dramatically, happens for the calculation using KB3G, although the strength integrated up to 7.5 MeV reproduces the experimental value quite well. The summed B(GT) up to $E_{x}=7.5$ MeV (a total of 48 states) for the KB3G interaction is $\sum B({\rm GT})_{\textrm{\scriptsize KB3G}}=2.02$ (a further 10\% of that value is located at energies up to 10.3 MeV) compared to the experimental value of $1.95\pm 0.14$ up to that excitation energy. The summed strength up to $E_{x}=7.5$ MeV with the GXPF1A interaction is $\sum B({\rm GT})_{\textrm{\scriptsize GXPF1A}}=2.65$. A further 8\% of that value is located at higher excitation energies, fragmented over many weak states.

A possible reason for the discrepancy between the data and the theory is a contribution from the $g_{9/2}$ orbit. In Ref. \cite{GAL98}, the observation of a collective rotational band in $^{64}$Zn, extending from spins 12$\hbar$ to $24\hbar$ with the band head at $\sim 8$ MeV was reported. That observation was explained by a model in which a hole was created in the proton $f_{7/2}$ orbit associated with triaxial deformation. One proton and two neutrons then resided in the $g_{9/2}$ orbit \cite{GAL98,RAG96}. Hence, at higher excitation energies one can expect contributions from the $g_{9/2}$ orbit, which are not included in the shell-model calculations performed for this work to calculate the GT strength distributions. At the excitation energies of relevance here, the influence of the $g_{9/2}$ orbit is not so clear. Theoretical studies of shape coexistence \cite{KAN04} indicate that the effects of the $g_{9/2}$ are small in Zn nuclei and in a recent study of the structure of $^{64}$Ge \cite{STA07}, $E2$ transition rates were well-described using the GXPF1A interaction. In summary, the intrusion of the $g_{9/2}$ shell likely plays a role, but not necessarily  at $E_{x}(^{64})$Cu$<6$ MeV. The strength distribution is also affected by the differences in deformation of $^{64}$Zn and $^{64}$Cu, which complicates matters further.

\section{\label{sec:ec}Application to Electron Capture Rates}

\begin{figure}
\begin{center}
\includegraphics[width=0.45\textwidth]{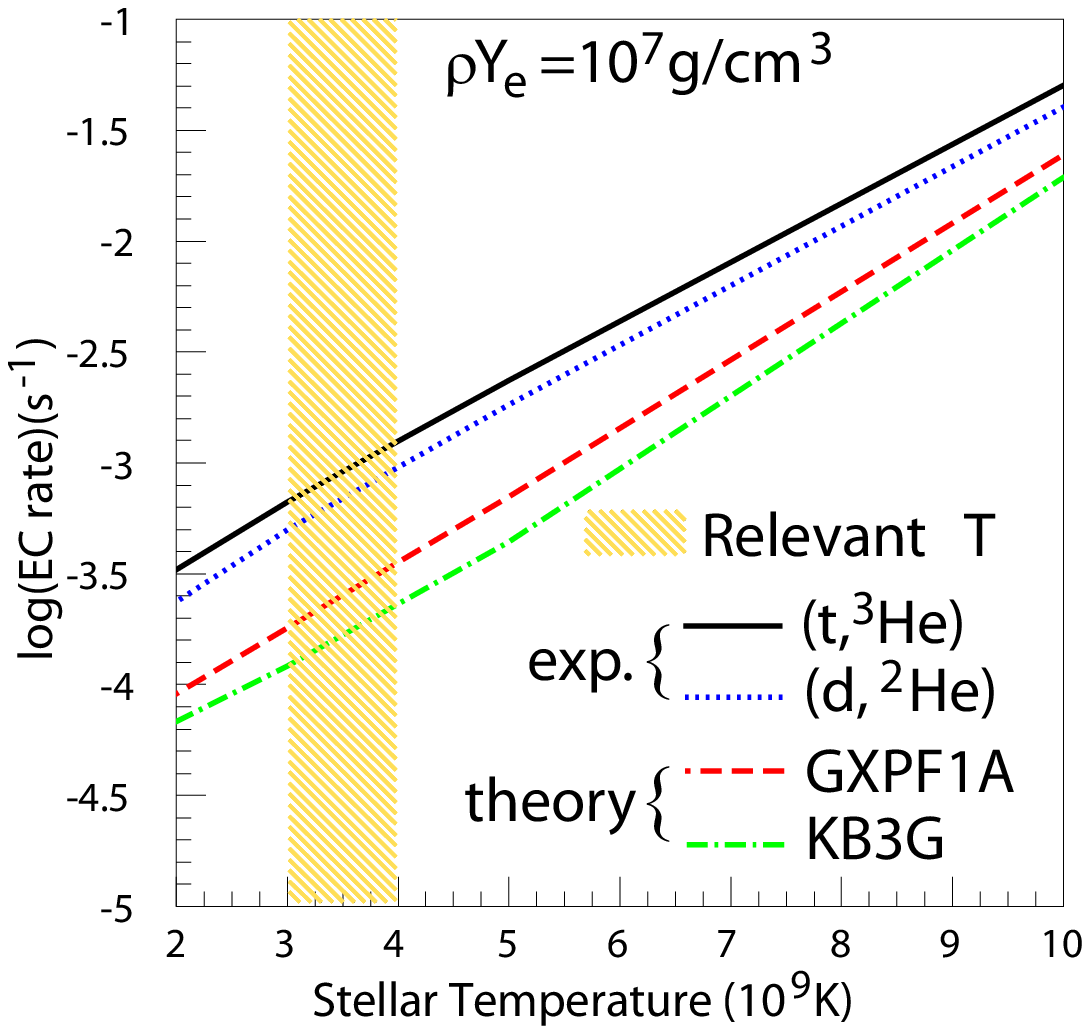}
\caption{(color online) Electron-capture rate on $^{64}$Zn, plotted as a function of stellar temperature, at an electron density of 10$^{7}$ g/cm$^{3}$.  Rates based on GT strength from the ($t$,$^{3}$He) data (black, solid line), the ($d$,$^{2}$He) data (blue, dotted line), the shell-model calculations using GXPF1A (red, dashed line) and KB3G (green, dashed-dotted line) are shown. The yellow cross-hatched region marks the relevant temperature range at this density in the pre-collapse trajectory for a $15M_{\rm solar}$ mass star.\label{fig8}}
\end{center}
\end{figure}

\begin{figure}
\begin{center}
\includegraphics[width=0.45\textwidth]{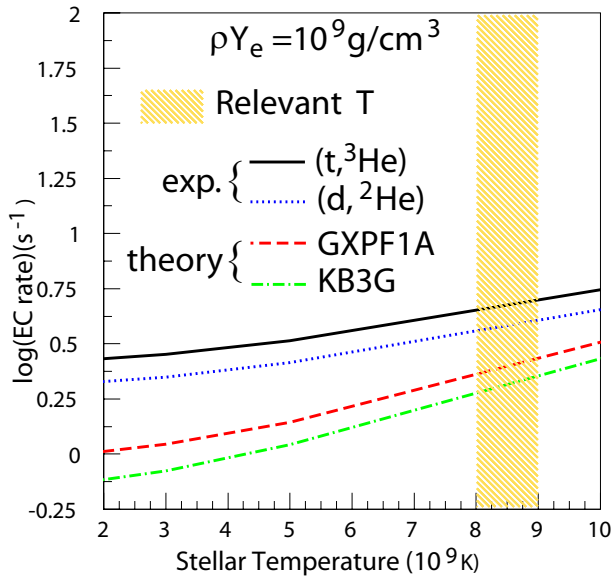}
\caption{(color online) Electron-capture rate on $^{64}$Zn, plotted as a function of stellar temperature, at an electron density of 10$^{9}$ g/cm$^{3}$.  Rates based on GT strength from the ($t$,$^{3}$He) data (black, solid line), the ($d$,$^{2}$He) data (blue, dotted line), the shell-model calculations using GXPF1A (red, dashed line) and KB3G (green, dashed-dotted line) are shown. The yellow cross-hatched region marks the relevant temperature range at this density in the pre-collapse trajectory for a $15M_{\rm solar}$ mass star.\label{fig9}}
\end{center}
\end{figure}

Calculations of the stellar EC rates onto the $^{64}$Zn groundstates were carried out following the procedures of Refs. \cite{full1980,full1982a,full1982b,full1985} using a code by Gupta {\it et al.} \cite{bece2006}. The capture rate was calculated over a wide grid of temperatures ($0.01\times 10^{9} \leq T \leq 100\times 10^{9} {\rm K}$) and electron densities ($10^{1} \leq \rho Y_{\rm e} \leq 10^{14} {\rm g/cm^{3}}$) which should cover nearly all scenarios relevant to nuclear astrophysics. Capture-rate calculations presented here are for two of the above electron densities; at $\rho Y_{\rm e}=10^{7}{\rm g/cm^{3}}$ and $\rho Y_{\rm e}=10^{9}{\rm g/cm^{3}}$, chosen for their relevance along the core-collapse supernova trajectory. Electron densities from $\rho Y_{\rm e}=10^{7}{\rm g/cm^{3}}$ to $\rho Y_{\rm e}=10^{9}{\rm g/cm^{3}}$ correspond to the conditions present in the core from the silicon-burning phase of the pre-collapse progenitor star to the conditions present at about 0.5 seconds before core-bounce, respectively. This time interval during the pre-bounce evolution is where electron-captures on nuclei have potentially the largest influence on the post-bounce trajectory \cite{hix2003,lang2004}.

Fig. \ref{fig8} shows the capture rate as a function of stellar temperature at $\rho Y_{\rm e}=10^{7}{\rm g/cm^{3}}$. The cross-hatched region indicates the interval of relevant temperatures for a 15 solar mass, main-sequence progenitor near the silicon burning phase. Capture-rate curves are plotted for the B(GT) extracted from the present experiment, that resulting from the B(GT) values published by Grewe \textit{et al.} \cite{grew2008}, and those resulting from B(GT) calculated using GXPF1A and KB3G shell-model interactions.

The electron Fermi energies for the density of $\rho Y_{\rm e}=10^{7}{\rm g/cm^{3}}$ is 0.7 MeV and the EC rates are thus very sensitive to the GT strength distribution at low excitation energies at this density. The capture rate based on the two measured B(GT) distributions differs by about 10\% in the window of relevant temperature. This is due to slight differences in the experimentally extracted GT strength distributions and the small discrepancy in the overall normalization discussed above. The level of disagreement between the rates based on the ($t$,$^{3}$He) and ($d$,$^{2}$He) data thus serves as a reasonable estimate of the systematic error in the capture rate determination from experiment. The comparative lack of strength in the first MeV of excitation energy predicted by both shell model calculations results in a severe underestimation of the capture rate. In Fig. \ref{fig8}, the capture rate using the GXPF1A interaction is lower by a factor of 3.5 and the KB3G result is lower by a factor of 5, relative to the rate based on the strengths extracted from the $^{64}$Zn($t$,$^{3}$He) experiment.

These differences between experiment and theory persist at higher densities, as shown in Fig. \ref{fig9}. There, the two measured capture rates still differ by about 10\%; the GXPF1A rate remains lower by a factor of 3.5, and the KB3G rate by a factor of 5 compared to the rate based on the present experimental data. This is in spite of the improved performance at increased Fermi energy (4.7 MeV at $\rho Y_{\rm e}=10^{9}{\rm g/cm^{3}}$) and higher excitation energies demonstrated by the GXPF1A calculation in Fig. \ref{fig4}. The reason for this, as pointed out by FFN \cite{full1980,full1982a}, is that while the branching ratio for capture to a given final state is proportional to its B(GT), the phase space available for electrons with an energy sufficient to excite levels at higher excitation energy in the daughter nucleus shrinks much more rapidly. Therefore, the total capture rate at the densities and temperatures studied here depends sensitively on the GT distribution at low excitation energies in $^{64}$Cu.

\section{\label{sec:conclu}Conclusions}

In summary, the GT strengths for  transitions from $^{64}$Zn to $^{64}$Cu have been extracted using the $^{64}$Zn($t$,$^{3}$He) CE reaction at 115 MeV/nucleon. The extraction procedure was based on the empirical proportionality between the cross section in the limit of vanishing momentum transfer and B(GT) derived for the ($^{3}$He,$t$) reaction. Systematic errors in the extraction due to action of the tensor operator in the effective NN interaction have been investigated. It breaks the proportionality by 10-20\% and is strongest for weaker GT transitions.

In spite of the different method employed to extract the B(GT), the experimentally extracted GT-strength distribution was, within error margins, consistent with that obtained from a $^{64}$Zn($d$,$^2$He) experiment \cite{grew2008}. The experimental GT strength distribution has also been compared to shell-model calculations using the \textsc{NuShellX} code employing the GXPF1A and KB3G effective interactions. Both sets of calculations fail to reproduce the data, although GXPF1A describes the strength distribution somewhat better than KB3G.

Since the shell-model calculations underestimate the GT strength at low excitation energies, deduced electron-capture rates on the $^{64}$Zn groundstate from the theoretical strength distributions using the GXPF1A (KB3G) interaction are too small compared to those deduced from the experimental distribution by a factor of 3.5 (5). The results indicate that further refinement of the interactions are required. It can furthermore be concluded that additional data is required to test the theoretical calculations and to check if the deviations are systematic in the upper $pf$-shell. An underestimation of the electron-capture rate on a single nucleus is not likely to strongly affect the impact of electron capture on astrophysical scenario, but systematic underestimation of electron-capture rates in a region of the nuclear chart can have significant consequences.

\begin{acknowledgments}
We thank the NSCL staff for their efforts in running this experiment and the CCF and A1900 operators in particular for the development of the high-intensity primary $^{16}$O and secondary triton beams. We also thank K.-L. Kratz, S. Gupta, P. M{\"o}ller, C. Caesar and V. Zelevinsky for discussions concerning the results.
This work was supported by the US-NSF (PHY0216783 (JINA), PHY0555366 and PHY-0758099) and US-DOE (DE-FC02-09ER41584).
\end{acknowledgments}

\bibliography{prc}

\end{document}